\documentstyle[12pt]{article}
\normalsize

\newcounter{sxn}

\newcounter{axn}

\def\br{\begin{thebibliography}{abc}}
\def\er{\end{thebibliography}}

\date{}

\begin{document}
\bibliographystyle{unsrt}
\footskip 1.0cm
\thispagestyle{empty}
\begin{flushright}
UMN-TH-1028/92\\
TPI-MINN-92/20-T\\
October 1992\\
\end{flushright}
\vspace*{6mm}
\centerline {\Large ON THE PRODUCTION OF FLUX VORTICES AND}
\vspace*{3mm}
\centerline {\Large MAGNETIC MONOPOLES IN PHASE TRANSITIONS}
\vspace*{8mm}
\centerline {\large Serge Rudaz}
\vspace*{5mm}
\centerline {\it School of Physics and Astronomy, University of Minnesota}
\centerline {\it Minneapolis, Minnesota 55455, USA}
\vspace*{8mm}
\centerline {\large Ajit Mohan Srivastava \footnote {\small Present
address: Institute for Theoretical Physics, University of
California, Santa Barbara, CA 93106, USA.} }
\vspace*{5mm}
\centerline {\it Theoretical Physics Institute, University of Minnesota}
\centerline {\it Minneapolis, Minnesota 55455, USA}

\vspace*{5mm}

\baselineskip=18pt

\centerline {\bf ABSTRACT}
\vspace*{3mm}

 We examine the basic assumptions underlying a scenario due to
Kibble that is widely used to estimate the production of topological defects.
We argue that one of the crucial assumptions, namely the geodesic rule,
although completely valid for global defects, becomes ill defined for
the case of gauged defects. We address the issues involved in
formulating a suitable geodesic rule for this case and argue that the
dynamics plays an important role in the production of gauge defects.

\newpage

\newcommand{\be}{\begin{equation}}
\newcommand{\ee}{\end{equation}}

\baselineskip=18pt
\setcounter{page}{2}

\vskip .1in
\centerline {\Large { 1. Introduction}}
\vskip .1in

  The existence of superheavy magnetic monopoles [1] and other types of
topologically non-trivial field configurations is an almost inevitable
consequence of unified gauge theories of elementary particle interactions.
A classical argument due to Kibble [2] (hereinafter referred to as ``the
Kibble mechanism") combined with a requirement of causality [3] leads to
an estimated lower limit on the number of magnetic monopoles produced
in a phase transition at Grand Unified Theory scale
(typically 10$^{16}$ GeV) in the early Universe
that is much too large to be compatible with the standard Big Bang cosmology.

  Various methods have been proposed to solve this ``cosmological monopole
problem", of which inflation is the best known [4]. In this letter, we
will reexamine the assumptions underlying the Kibble mechanism in some
detail.  These assumptions are, first, the existence of uncorrelated
regions of space in which the vacuum degrees of freedom of scalar fields
fluctuate randomly and second, what is usually called the geodesic rule,
namely, that in between any two such uncorrelated regions of space,
the scalar field traces the shortest path on the vacuum manifold.

We will argue in what follows that the geodesic rule in its conventional
form becomes ambiguous when applied
to the production of the topological defects in theories with
spontaneously broken (local) gauge symmetries.
This will lead us to suggest alternate physical criteria to modify
the Kibble argument for this case. The original Kibble argument
remains applicable without modification to the case of defects arising
from the spontaneous breaking of global symmetries.

\vskip .1in
\centerline {\Large { 2. Formation of Global Defects}}
\vskip .1in

 We start by examining the global case.
For concreteness we will consider the case of strings
arising due to the spontaneous breaking of a global U(1)
symmetry in a first order phase transition where the transition proceeds
through nucleation of vacuum bubbles [5]. We consider first order transition
as in this case certain assumptions such as the existence of uncorrelated
regions can be clearly formulated. The considerations are easily
extended later to second order phase transition and to other topological
defects. Let us assume that bubbles of a critical size randomly nucleate
in space and collide with each other as they continue to expand.
The strings form at the intersection of three (or more) bubbles if the
Higgs phases in the three bubbles have appropriate values: Note that string
production due to two bubbles (where the Kibble argument is in any case not
applicable) has been shown to be extremely unlikely by numerical
simulations for the 2+1 dimensional case [6]. Due to the minimization
of gradient energy, the Higgs phase in a given bubble will be
expected to be uniform [6].
The three bubbles here represent three uncorrelated regions where the
Higgs phase varies randomly from one region to the other. The existence of
such uncorrelated regions is one of the crucial elements in the
Kibble mechanism and, at least for the global case,
is well justified by physical considerations.
However, this alone is not enough to determine the production of
strings and one needs to make the following additional crucial assumption
regarding  the spatial variation of the Higgs phase.

 Consider the collision of two bubbles in the above picture.
As the bubbles coalesce, some intermediate value of the
Higgs phase will arise in the region where the bubbles
coalesced. It is immediately clear that the gradient energy will force
the Higgs phase to smoothly interpolate between the two bubbles
such that it traces the shortest path on U(1) (this is also verified in the
numerical simulations, see [6]). It is important to
realize that this intermediate value of the Higgs phase is completely
determined by the values of Higgs phase in the two
(initially separated) bubbles
and follows from the minimization of gradient energy. This is the
geodesic rule crucial to the application of the Kibble mechanism.

 These two ingredients are now enough to determine the string
production. As the three bubbles coalesce, one can trace
out a closed path in the region of the true vacuum formed by these bubbles.
If the Higgs phase changes by 2n$\pi$  (where n is a non-zero integer) around
this path, one can conclude that eventually a string must form inside.
The expected string number per correlation volume
within this scenario can be calculated to be 1/4 for 2 space dimensions
and about 0.88 for 3 space dimensions. Again we emphasize the
important aspects of this approach: First,
the intermediate values of Higgs phase were determined by the local
physics following energy considerations, and second, the signal for the
formation of string came from the outer region by considering a large
closed path along which the Higgs phase changed by 2n$\pi$.

All of what has been said so far is valid for the case of second order
phase transition as well. For example, the bubbles can represent
various correlation volumes (or, as a limiting case, horizon volumes)
and then the Higgs phase in the intermediate region of any
two such volumes will be completely determined by the gradient energy
considerations which will again enforce the geodesic rule. Also,
all these considerations  can be clearly
generalized for other global defects  such as monopoles, by
considering appropriate number of bubbles etc. [7].
We remark here that, due to strong attractive forces, global monopole -
antimonopole pairs annihilate very efficiently and their number density
is quite consistent with observations [8]

\vskip .1in
\centerline {\Large {3. Gauged Defects and the Geodesic Rule}}
\vskip .1in

 Let us now consider the case of gauged strings. The Kibble argument
can only be formulated once a non-unitary gauge choice has been made.
Clearly, the fact that the phase of a Higgs field is not a gauge-invariant
quantity should be of no consequence, provided the argument leading to an
estimate of a gauge-invariant quantity like the number of topological defects,
is properly carried out.
Consider the case when a U(1) local gauge symmetry is
spontaneously broken, and again at first that the phase transition is of
first order. The first assumption of the Kibble mechanism
concerning the existence of uncorrelated regions still seems reasonable
as this is consistent with the spirit of local physics. Thus for the
above case of three colliding bubbles we will assume that the
value of the Higgs phase varies randomly from one bubble to another
(which essentially amounts to a specific non-unitary gauge choice).
As the inside of these bubbles are superconducting regions,
the magnetic field must be zero inside and the vector potential there
will be pure gauge.

 Now let us examine the collision of two bubbles; further, again assume that
the Higgs phase is uniform within a given bubble, which can be achieved
by a suitable gauge choice and the condition that the covariant derivative
$D_\mu \phi$ = 0 inside a bubble.  In the global case, this
uniformity followed from the local minimization of $\partial_\mu \phi$
(where $\phi$ is the Higgs field) due to considerations of
gradient energy which also determined the value of the Higgs
phase in the intermediate region as the bubbles coalesced.
However, in the presence of gauge fields, the gradient energy to be
minimized involves $D_\mu \phi$, rather
than just $\partial_\mu \phi$: Clearly, this is not sufficient to
uniquely determine the value of the Higgs phase itself that arises as the
bubbles coalesce, let alone to justify the statement that this quantity traces
the shortest path on the vacuum manifold.

Here we may mention that there have
been several numerical studies of the formation of gauged vortices [9,10] where
it has been found that large number of vortices are produced. However, in
these studies, the assumption of geodesic rule in one form or another is always
present. For example, in [9], random values of the Higgs phase are assumed from
one horizon volume to another while the values of the Higgs phase for points
within a horizon volume are determined by using smooth interpolation.
Similarly, in [10], initial vortices are identified by implicitly using the
geodesic rule for the Higgs phase. Clearly,
in such schemes the gauge field becomes irrelevant for determining the
initial number density of defects (even though it affects the evolution
of strings, see [10] for example) and one would expect the same initial number
of defects as in the global case. It is this application of geodesic rule
for the Higgs phase which is the subject of our analysis.

One may argue that in the absence
of any magnetic fields one can choose a gauge such that the vector potential
$A_\mu$ is zero everywhere so that $D_\mu \phi$ reduces to
$\partial_\mu \phi$ and  one may recover the geodesic rule for the Higgs phase.
One can certainly choose such a gauge before the bubbles collide.
But then one can not guarantee that $A_\mu$ will remain zero at the time
when the bubbles coalesce. On the other hand, if one wanted to make
such a gauge choice (so that $A_\mu$ = 0) at the time when the bubbles
collide, then the assumption of different values of $\theta$ in the
two bubbles may prohibit one from doing so. For colliding bubbles, such a
gauge choice seems possible  only for the case when $D_\mu \phi$ is identically
zero at the junction (which would mean that with the gauge choice in which
$A_\mu$ = 0, $\theta$ will be uniform in the two bubbles). This is because
if $D_\mu \phi$ is non-zero then there will be currents at the junction
when the bubbles touch leading to magnetic fields and hence non-zero $A_\mu$.
The geodesic rule is relevant only when the bubbles touch each other and as we
have just argued, at that moment it is generally not possible to choose a gauge
in which $A_\mu$ is zero. This again shows that the geodesic rule for
$\theta$ can not be applied here (except for the trivial case when
$\theta$ is uniform in the two bubbles, with $A_\mu$ = 0).

As far as the spatial variation
of the Higgs phase alone is concerned, any possible continuous variation
of the Higgs phase is allowed on an
open path in  the region intermediate to the two bubbles,
all such possibilities being related to each other
by gauge transformations and hence physically
equivalent. {\it We would like to emphasize here that we are not saying
that any variation of Higgs phase can be gauged away}. In the broken
phase the net change in the Higgs phase around a closed path
is gauge invariant. Similarly the integral of the vector potential
around a closed path (holonomy) is gauge invariant. What we are
saying here is that the change in the Higgs phase (or the change in the vector
potential in the superconducting region) from one point to another different
point is completely
gauge dependent. As the geodesic rule concerns the variation of
Higgs phase from one point to another different point, we conclude
that for gauge theories the geodesic rule for $\theta$ alone becomes
unjustified. Without a criterion like the geodesic rule, one can not specify
the configuration of $\theta$ on a closed path and hence can not determine
whether the path encloses any string or not.

\vskip .1in
\centerline {\Large {4. Possible Alternatives to the Use of the Geodesic Rule}}
\vskip .1in

 Given the above discussion, the immediate natural thing to do would be  to
concentrate on $D_\mu \phi$ (or more appropriately, on the physically well
defined gauge invariant quantity $\phi^\dagger D_\mu \phi$) in order to
establish a geodesic like criteria. We now analyze this possibility.
As the energy minimization will imply that $D_\mu \phi = 0$
inside the two colliding  bubbles (at least initially), we may work in
a gauge where the Higgs phase is uniform inside each bubble with
values $\theta_1$ and $\theta_2$ respectively. The vector potentials
are then zero inside each of the two bubbles. In the absence of any
magnetic fields, one can work in a gauge such that $A_\mu$ is zero initially
in the entire region (assuming that the same gauge choice still allows
for random variation of $\theta$ in the two bubbles). Note that, as we
discussed above, this is possible in general only when the bubbles are
separated so that we can assume that there are no currents anywhere.
Now as the bubbles are brought together, the question arises that what
will be the value of $D_\mu \phi$ at the junction of the bubbles.
There are two possibilities and we consider them in the following.

One possibility is that $A_\mu$ evolves in the intermediate
region so that so that $\int_1^2 {\vec A}.{\vec dl} = \theta_1 - \theta_2$
leading to $D_\mu \phi$ = 0 at the bubble junction.
Indeed, this is what one will expect if the bubbles collide very slowly
(for example when bubbles nucleate very rapidly
so that they expand very little, with small velocity, before colliding with
each other). However, as now a  non-zero value of $A_\mu$ has been
generated (say only inside the bubbles, near the collision region) one
should properly account for any associated magnetic field. If we consider
one junction at a time then it is clear that such a magnetic field will
arise initially in the form of a closed ring around the perimeter of
the collision area. Such a ring of magnetic field should quickly
shrink down and dissappear near the mid point of the collision, again
if the dynamics is slow. For fast dynamics it is conceivable that such
rings of magnetic fields will form at each of the junctions which will
then quickly combine to give one vortex. It seems interesting to investigate
the structure of this vortex comprising of three rings (at least initially)
which should result in such type of formation process. However, for slow
enough dynamics, each of these rings should shrink down at the associated
junctions and no vortex formation should result. When these rings shrink
down, wavefronts of $\theta$ and $A_\mu$ will emanate from that region
to make the field distribution consistent with the absence of the vortex.
For example, if the values of $\theta$ in the two bubbles were (say) 0
and $2 \pi/3$ respectively, then after the magnetic field dissipates in
the above manner it is completely possible that $\theta$ in the middle
winds around the longer path on U(1) rather than having value $2 \pi/6$
in that region. Which direction the variation of $\theta$
should choose on U(1) should be governed
by the condition whether there is a vortex formation or not. Note that
here we are reversing the standard sequence of arguments as applied to
the case of global defects. For the global case, the
winding of $\theta$ was first deduced and then vortex formation was
concluded. Here we are saying that since open paths of $\theta$
on U(1) do not have any physical meaning, they can not govern the production
of vortex but rather the production of vortex should decide the path of
$\theta$ on U(1).

 The above possibility was when the bubble collision is very slow. If this
is not true then it is possible that $A_\mu$ remains zero until the walls
collide  (for example when bubble walls are extremely relativistic as
in ref. [11]) and there may be non-zero $D_\mu \phi$ generated at the junction.
This will then lead to transient currents at the junction which could add up
and lead to the vortex formation.

 The situation here
is similar to that of the Josephson effect [12].
The two bubbles can thus be thought of as two pieces of superconductors
with the region of false vacuum between the bubbles representing the
Josephson junction. The net current across the Josephson junction is
determined by external conditions which can be, either the presence of a
potential difference across the junction, or the presence of a
magnetic field at the junction, or the fact that there is supercurrent
flowing through the circuit of which the Josephson junction is a part [12].
However, even if there are no external influences, there may still be a
transient current across the junction which, for a single junction, will
eventually relax to zero but may possibly add up in a closed array of more
than two junctions. For example one may consider three Josephson junctions,
with the value of $D_\mu \phi$ always having same sign across each of the
junctions (which can be arranged by choosing $A_\mu$ = 0 and monotonically
increasing value of the Higgs phase as one goes around the shaded region),
then the transient currents may add up to give a net circulating current
with corresponding formation of vortex. This indeed suggests the
intriguing possibility of carrying out such an experiment (for example in
Josephson junction arrays) where the probability
of spontaneous generation of magnetic flux can be experimentally determined.

These arguments can be extended to the case of
a second order phase transition. Thus
the bubbles can be taken to represent different correlation
volumes as now the whole region is in the same phase (depending
on the temperature). The essentials of our argument are still
the same. As far as generation of non-zero $D_\mu \phi$
is concerned, in this case it should be even more natural to expect
(due to more homogeneous nature of the phase transition)
that, as different uncorrelated regions come into contact, the vector
potential will develop such that $D_\mu \phi$ relaxes to zero. One has
to then consider the magnetic fields generated by the evolution of
$A_\mu$. Though it seems likely that here as well the magnetic field
may arise in the form of a ring (at the junction of the two domains)
which may then shrink down. This is especially likely when the vector
boson in the broken phase is very heavy so that the time scale associated
with it is very short. Extensions of our arguments to the case of
magnetic monopoles can be carried out straightforwardly by considering
the collision of four bubbles and seeing whether non-zero values of
$D_\mu \phi$ can arise at the junctions. If it does turn out that for
certain types of dynamics the production of gauge defects through the
Kibble mechanism is suppressed then a competitive mechanism for their
production may be through thermal fluctuations [13].

\vskip .1in
\centerline {\Large { 5. Conclusions}}
\vskip .1in

  We conclude by emphasizing the main points of this paper. What we have
argued is that it is not appropriate to treat gauge defects in exactly the
same way as global defects. The analysis of the formation of gauge defects
requires much more detailed considerations and it seems possible that the
dynamics plays a very crucial role in this case. This is because
a simple criterion like the geodesic rule for the Higgs phase is not available
here which could be used to complete the specification of field configuration
given the values of Higgs phases in two (separated) uncorrelated regions.
One therefore can not trust the conventional estimates of the gauged monopole
(or string) number density which are based on the use of the geodesic rule
for $\theta$ alone. A more thorough analysis of the dynamics at
the phase transition is essential in the case of gauge theories, and the
mainly topological arguments appropriate to the global case are not sufficient.
It is possible that transient currents that fail to dissipate given
sufficiently rapid dynamics could act coherently and lead to the creation of
the gauged topological defects. On the other hand it also seems possible that
the magnetic fields may arise only in the form of rings around individual
junctions of colliding bubbles which then shrink down.
These issues can be settled by extending the study of [11]
by considering three-bubble collisions with varying field configurations
and bubble wall velocities. We are presently investigating this.
As our arguments suggest that there may be a difference in the production
of global and gauge defects, it is interesting to see if
the formation of flux tubes in type II superconductors can be experimentally
tested along the same lines as the test for the global case in liquid
crystal systems [14]. We are presently investigating this and the
results will be presented in a subsequent paper [15].

\vskip .1in
\centerline {\Large { Acknowledgements}}
\vskip .1in

 We thank Y. Hosotani, A.P. Balachandran, A. Casher, C. Dasgupta, A. Goldman,
Fong Liu, Joe Polchinski, C. Rosenzweig and Michael Stone for useful
discussions. This work was supported by the U.S.
Department of Energy under contract number DE-AC02-83ER40105.
The work of A.M.S. was also supported by the Theoretical
Physics Institute at the University of Minnesota.

\newpage

\baselineskip=18pt
\centerline{\Large {References}}
\begin{enumerate}

\item G. 't Hooft, Nucl. Phys. B79, 276(1974); A.M. Polyakov, Zh. Eksp. Teor.
Fiz. Pis. Red. 20, 430 (1974) [JETP Letters 20, 194(1974)].

\item T.W.B. Kibble, J. Phys. A9, 1387(1976).

\item M. B. Einhorn, D. L. Stein, and D. Toussaint,
Phys. Rev. D21, 3295(1980).

\item For a review, see K. A. Olive, Phys. Rep. 190, 307(1990).

\item M.B. Voloshin, I.Yu. Kobzarev, and L.B. Okun, Yad. Fiz. 20, 1229(1974)
[Sov. J. Nucl. Phys. 20, 644(1975)]; S. Coleman, Phys. Rev. D15, 2929(1977);
A.D. Linde, Nucl. Phys. B216, 421(1983).

\item A. M. Srivastava, Phys. Rev. D45, R3304(1992); Phys. Rev. D46,
1353(1992).

\item For a review, see A. Vilenkin, Phys. Rep. 121, 263(1985).

\item D. P. Bennett and S. H. Rhie, Phys. Rev. Lett. 65, 1709(1990).

\item J. Ye and R. H. Brandenberger, Nucl. Phys. B346, 149(1990).

\item F. Liu, M. Mondello and N. Goldenfeld, Phys. Rev. Lett. 66, 3071(1991).

\item S.W. Hawking, I.G. Moss, and J.M. Stewart, Phys. Rev. D 26, 2681(1982).

\item B.D. Josephson, Adv. Phys. 14, 419(1965); P.W. Anderson, Progress
in Low Temperature Physics, Vol. V, 1 (1967).

\item F. A. Bais and S. Rudaz, Nucl. Phys. B170, 507(1980).

\item M.J. Bowick, L. Chandar, E.A. Schiff and A.M. Srivastava,
Syracuse/Minnesota preprint SU-HEP-4241-512/TPI-MINN-92/35-T, (August,1992).

\item S. Rudaz, A.M. Srivastava and S. Varma, (in preparation).

\end{enumerate}

\end{document}